\documentclass[aps,pra,twocolumn,nobibnotes,superscriptaddress,nofootinbib,10pt]{revtex4-1} 
\usepackage[latin1]{inputenc}
\usepackage{amsmath,amssymb}
\usepackage{mathrsfs} 
\usepackage{graphicx,color,colortbl}
\usepackage{rotating,array,tabularx,booktabs}
\usepackage{varwidth,xcolor}
\usepackage{placeins}
\usepackage{siunitx}
\usepackage[normalem]{ulem}% Fuer Durchstreichungen

\DeclareGraphicsExtensions{.pdf,.PDF,.png,.PNG}
\allowdisplaybreaks

\newcommand{\dif}{\mathrm{d}}%
\newcommand{\ZT}[1]{\textquotedblleft#1\textquotedblright}%
\newcommand{\uu}{\hat{n}}%
\newcommand{\ww}{\vec{u}}%
\newcommand{\ws}{u}%

\newcommand{\NichtAnwendbar}{---}%

\begin{document}

\title{Ultrasound-propelled nano- and microspinners}

\author{Johannes Vo\ss{}}
\affiliation{Institut f\"ur Theoretische Physik, Center for Soft Nanoscience, Universit\"at M\"unster, 48149 M\"unster, Germany}

\author{Raphael Wittkowski}
\email[Corresponding author: ]{raphael.wittkowski@uni-muenster.de}
\affiliation{Institut f\"ur Theoretische Physik, Center for Soft Nanoscience, Universit\"at M\"unster, 48149 M\"unster, Germany}

\begin{abstract}
We study nonhelical nano- and microparticles that, through a particular shape, rotate when they are exposed to ultrasound. Employing acoustofluidic computer simulations, we investigate the flow field that is generated around these particles in the presence of a planar traveling ultrasound wave as well as the resulting propulsion force and torque of the particles. We study how the flow field and the propulsion force and torque depend on the particles' orientation relative to the propagation direction of the ultrasound wave. Furthermore, we show that the orientation-averaged propulsion force vanishes whereas the orientation-averaged propulsion torque is nonzero. Thus, we reveal that these particles can constitute nano- and microspinners that persistently rotate in isotropic ultrasound. 
\begin{figure}[htb]
\centering
\fbox{\includegraphics[width=8cm]{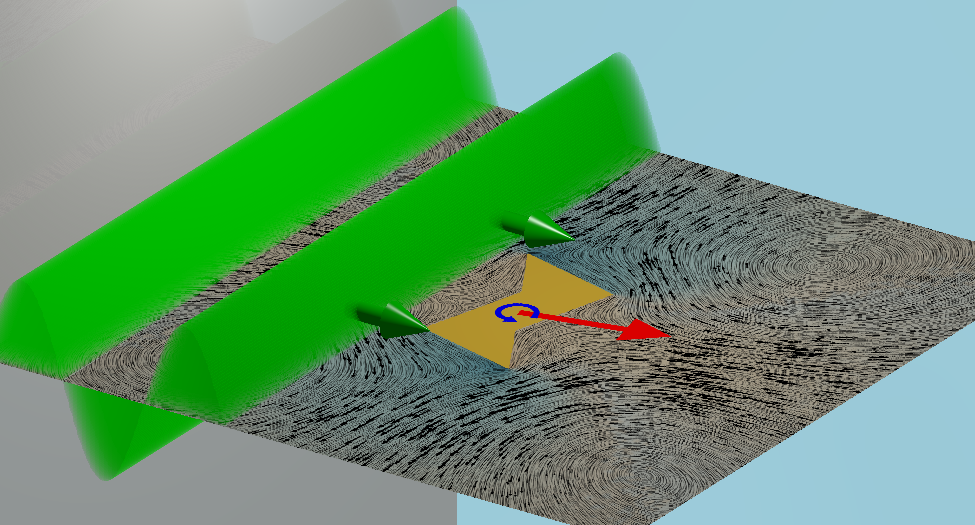}}%
\end{figure}
\end{abstract}
\maketitle

\section{Introduction}
Ultrasound-propelled nano- and microparticles, discovered in 2012 \cite{WangCHM2012}, are artificial particles that become motile when they are exposed to ultrasound. 
During the past decade, they have received large scientific interest and developed into an important and still rapidly growing field of research \cite{WangCHM2012,GarciaGradillaEtAl2013,AhmedEtAl2013,NadalL2014,WuEtAl2014,WangLMAHM2014,GarciaGradillaSSKYWGW2014,BalkEtAl2014,AhmedGFM2014,WangDZSSM2015,EstebanFernandezdeAvilaMSLRCVMGZW2015,Kiristi2015,WuEtAl2015a,WuEtAl2015b,RaoLMZCW2015,ZhouZWW2017,KimGLZF2016,EstebanEtAl2016,SotoWGGGLKACW2016,AhmedWBGHM2016,KaynakONNLCH2016,UygunEtAl2017,EstebanFernandezEtAl2017,RenZMXHM2017,CollisCS2017,ZhouYWDW2017,ChenEtAl2018,HansenEtAl2018,SabrinaTABdlCMB2018,AhmedBJPDN2016,Zhou2018,WangGWSGXH2018,EstebanEtAl2018,BhuyanDBSGB2019,LuSZWPL2019,TangEtAl2019,QualliotineEtAl2019,GaoLWWXH2019,KaynakONLCH2017,RenEtAl2019,AghakhaniYWS2020,McneillSWOLNM2021,LiuR2020,VossW2020,ValdezLOESSWG2020,DumyJMBGMHA2020,VossW2021,VossW2022acoustica,VossW2022orientation,VossW2022acousticb}. 
Besides the general growing interest in artificial motile nano- and microparticles \cite{BechingerdLLRVV2016,Venugopalan2020,FernandezRMHSS2020,YangEtAl2020,FratzlFKS2021} (so-called \ZT{active colloidal particles} \cite{ZhangLGG2017,SabrinaTABdlCMB2018,DouB2019}), the investigation and further development of acoustically propelled particles benefited from their outstanding properties and application relevance. 
For example, acoustically propelled particles can, via an ultrasound field, easily and persistently be supplied with energy \cite{XuGXZW2017,WangLWXLGM2019} and their propulsion mechanism works in various types of fluids including biofluids \cite{GarciaGradillaEtAl2013,WuEtAl2014,WangLMAHM2014,EstebanEtAl2018,GaoLWWXH2019,WangGZLH2020,WuEtAl2015a,EstebanFernandezdeAvilaMSLRCVMGZW2015,EstebanEtAl2016,EstebanFernandezEtAl2017,UygunEtAl2017,HansenEtAl2018,QualliotineEtAl2019}. 
Furthermore, acoustic propulsion has been found to be biocompatible \cite{WangLWXLGM2019,OuEtAl2020}. 
These advantages, compared to most of the other types of artificial motile particles that have been developed so far \cite{EstebanFernandezdeAvilaALGZW2018,SafdarKJ2018,PengTW2017,KaganBCCEEW2012,XuanSGWDH2018,XuCLFPLK2019,FernandezRMHSS2020}, make acoustically propelled particles relevant for a number of important potential future applications, such as targeted drug delivery \cite{LuoFWG2018,ErkocYCYAS2019,JinYDWWZ2021,OuEtAl2021}. 

Up to now, the investigation of ultrasound-propelled nano- and microparticles has already resulted in many insights into their properties and in improvements of their design \cite{VossW2021,LiMMOP2021}.
For example, particles with various shapes have been investigated and the particle shape has been found to have a large influence on the particles' acoustic propulsion 
\cite{CollisCS2017,VossW2020,VossW2021,AhmedBJPDN2016,AhmedWBGHM2016,SotoWGGGLKACW2016,LiMMOP2021}. 
Previous studies covered particles with a rigid shape \cite{WangCHM2012,GarciaGradillaEtAl2013,AhmedEtAl2013,NadalL2014,BalkEtAl2014,AhmedGFM2014,GarciaGradillaSSKYWGW2014,WangDZSSM2015,EstebanFernandezdeAvilaMSLRCVMGZW2015,Kiristi2015,SotoWGGGLKACW2016,EstebanEtAl2016,AhmedWBGHM2016,UygunEtAl2017,CollisCS2017,HansenEtAl2018,EstebanEtAl2018,SabrinaTABdlCMB2018,TangEtAl2019,ZhouZWW2017,VossW2020,ValdezLOESSWG2020,Zhou2018,RenWM2018,DumyJMBGMHA2020,VossW2021,VossW2022acoustica,VossW2022orientation,VossW2022acousticb,McneillSWOLNM2021} as well as particles with a deformable shape \cite{KaganBCCEEW2012,AhmedLNLSMCH2015,AhmedBJPDN2016,KaynakONLCH2017,ZhouYWDW2017,WangGWSGXH2018,RenEtAl2019,AghakhaniYWS2020,LiuR2020}. 
The deformable particles can achieve quite large propulsion speeds but are more difficult to fabricate than the rigid ones that are thus more likely to be used in future applications \cite{VossW2020,VossW2021,VossW2022orientation,LiMMOP2021}.
In the case of rigid shapes, bullet-shaped  \cite{WangCHM2012,GarciaGradillaEtAl2013,AhmedEtAl2013,AhmedGFM2014,BalkEtAl2014,WangLMAHM2014,EstebanFernandezdeAvilaMSLRCVMGZW2015,Kiristi2015,AhmedWBGHM2016,ZhouZWW2017,WangGWSGXH2018,ZhouYWDW2017,Zhou2018,DumyJMBGMHA2020,McneillSWOLNM2021}, bowl-shaped \cite{SotoWGGGLKACW2016,TangEtAl2019,VossW2020}, cone-shaped \cite{VossW2020,VossW2021,VossW2022acoustica,VossW2022orientation,VossW2022acousticb}, and gear-shaped \cite{KaynakONNLCH2016,SabrinaTABdlCMB2018} particles have been investigated. 
Most of these particles show translational propulsion when they are exposed to ultrasound, but there is also a small number of studies that considered particles with predominantly rotational propulsion, so-called nano- and microspinners \cite{WangCHM2012,BalkEtAl2014,AhmedGFM2014,AhmedLNLSMCH2015,KimGLZF2016,KaynakONNLCH2016,ZhouZWW2017,KaynakONLCH2017,SabrinaTABdlCMB2018,LiuR2020,MohantyEtAl2021}. 
However, much research is still needed until acoustically propelled nano- and microparticles can actually be applied in nanomedicine and other envisaged areas \cite{Venugopalan2020,LiMMOP2021}. 
For example, the past investigation of acoustically propelled particles has mainly focused on particles with translational propulsion, since they are important for future applications like targeted drug delivery, so there has been relatively little research on particles with rotational propulsion \cite{WangCHM2012,BalkEtAl2014,AhmedGFM2014,AhmedLNLSMCH2015,KimGLZF2016,KaynakONNLCH2016,ZhouZWW2017,KaynakONLCH2017,SabrinaTABdlCMB2018,LiuR2020,MohantyEtAl2021}. 
While the latter particles would not be a good choice for targeted drug delivery, they are important for other potential future applications. Nano- and microspinners could, e.g., be applied as nano- and micromixers to mechanically mix otherwise immiscible fluids on a microscopic level. When such mixers are suitably functionalized (e.g., by attaching surfactants to their surface), they can assemble at the interface between the fluids that shall be mixed and thus provide an opportunity for selective, interface-related mixing. Potential advantages of this type of mixing are minimization of the mechanical stress that is exerted on the fluids, which minimizes heating up of the fluids and can be advantageous in the case of thermally fragile substances, the option to control the amount of mixing in space and time by modulating the particles' propulsion, orientation, and position accordingly \cite{NitschkeW2021}, and unconventional pattern formation. 
It is, therefore, appropriate to place greater emphasis on the investigation of ultrasound-propelled nano- and microspinners.
Furthermore, as all of the few existing studies on such particles focus on particles in a planar \cite{WangCHM2012,BalkEtAl2014,AhmedGFM2014,ZhouZWW2017,SabrinaTABdlCMB2018,LiuR2020,KaynakONNLCH2016,KaynakONLCH2017} or circular \cite{MohantyEtAl2021} standing ultrasound wave, it is advisable to start investigating their behavior in other types of ultrasound fields, such as a traveling ultrasound wave and isotropic ultrasound, which are more relevant with respect to future applications of acoustically propelled particles \cite{VossW2020,VossW2021,VossW2022orientation,VossW2022acoustica,VossW2022acousticb,AhmedBJPDN2016}.  
In this manuscript, we, therefore, aim at advancing the investigation of ultrasound-propelled nano- and microspinners.  
For this purpose, we propose a new particle design, study the orientation-dependence of its propulsion in a planar traveling ultrasound wave, and show that it exhibits persistent rotational motion when exposed to isotropic ultrasound.   
For our investigation, we performed direct acoustofluidic computer simulations that are based on numerically solving the compressible Navier-Stokes equations.

\section{\label{methods}Methods}
Our methodology follows Ref.\ \cite{VossW2020}. We have adopted this methodology since it has been proven to be successful.

\subsection{Setup}
\begin{figure}[htb]
\centering
\includegraphics[width=\linewidth]{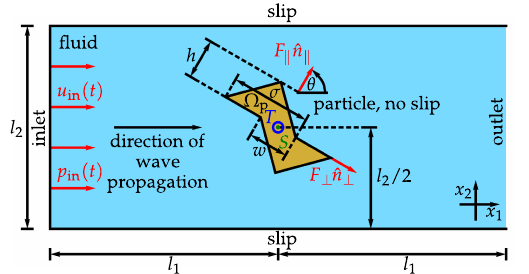}%
\caption{\label{fig:fig1}Setup for the simulations.}
\end{figure}

The simulated system is shown in Fig.\ \ref{fig:fig1}. 
It consists of a particle in a fluid-filled rectangular domain. 

The rectangular simulation domain has width $2l_1$ (aligned with the $x_1$-axis) and height $l_2=\SI{200}{\micro\metre}$ (aligned with the $x_2$-axis).
As the fluid, we choose water with a vanishing velocity field $\ww_0=\vec{0}\,\SI{}{\metre\,\second^{-1}}$ at time $t=0$, where the simulations start. 
The quiescent initial water is at standard temperature $T_0=\SI{293.15}{\kelvin}$ and standard pressure $p_0=\SI{101325}{\pascal}$. 

We place the particle's center of mass $\mathrm{S}$ in the center of the simulation domain.
The particle has a shape that can be obtained by joining two oppositely oriented triangular subparticles with diameter $\sigma=\SI{1}{\micro\metre}$ and height $h=\sigma/2$ bottom to bottom with an overlap of width $w=\sigma/2$. 
We define the orientation of the particle as the orientation of the symmetry axis of one of the triangular subparticles (see Fig.\ \ref{fig:fig1}). 
This orientation can be described by a unit vector $\uu_\parallel$.
An orientation perpendicular to this, which is parallel to the bottom edges of the triangular subparticles, is then given by another unit vector $\uu_\perp$.
Introducing a polar angle $\theta$ that is measured anticlockwise from the positive $x_1$-axis, the orientational unit vectors can be parameterized as 
\begin{align}
\uu_\parallel(\theta) &= (\cos(\theta),\sin(\theta))^\mathrm{T},\\
\uu_\perp(\theta) &= (\sin(\theta),-\cos(\theta))^\mathrm{T}.
\end{align}
In this work, we call $\theta$ the orientation of the particle. 
We vary the orientation from $\theta=0$, where $\uu_\parallel$ points into the direction of propagation of the ultrasound wave, to $\theta=\pi$, where the particle and the ultrasound wave point into opposite directions. 

At the left edge of the simulation domain, we prescribe inlet boundary conditions so that a planar traveling ultrasound wave enters the system and can propagate parallel to the $x_1$-axis.
We choose slip boundary conditions at the lower and upper edges of the simulation domain to avoid unwanted damping of the ultrasound wave there. 
For the right edge of the simulation domain, we prescribe outlet boundary conditions so that the wave can leave the system.
At the particle's surface $\partial\Omega_\mathrm{p}$, which confines the particle domain $\Omega_\mathrm{p}$, we prescribe no-slip boundary conditions to ensure a realistic interaction of the particle with the ultrasound wave. 

The ultrasound wave entering the simulation domain is prescribed through a time-dependent velocity $\ws_{\mathrm{in}}(t)=\Delta \ws\sin(2\pi f t)$ and pressure $p_{\mathrm{in}}(t)=\Delta p \sin(2\pi f t)$
at the inlet, where $t$ denotes time. 
$\Delta\ws=\Delta p/(\rho_0 c_\mathrm{f})$ is the flow velocity amplitude, and we choose $\Delta p=\SI{10}{\kilo\pascal}$ for the pressure amplitude and $f=\SI{1}{\MHz}$ for the frequency of the ultrasound wave. 
Furthermore, $\rho_0=\SI{998}{\kilogram\,\metre^{-3}}$ is the initial mass density of the fluid and $c_\mathrm{f}=\SI{1484}{\metre\,\second^{-1}}$ is its sound velocity. 
The ultrasound has thus the wavelength $\lambda=c_\mathrm{f}/f=\SI{1.484}{\milli \metre}$ and the acoustic energy density $E=\Delta p^2/(2 \rho_0 c_{\mathrm{f}}^2)=\SI{22.7}{\milli\joule\,\metre^{-3}}$. 
We choose the width of the simulation domain so that $l_1=\lambda/4$. 

When the ultrasound wave interacts with the particle, a propulsion force and a propulsion torque are exerted on the particle's center of mass $\mathrm{S}$. We are mainly interested in the stationary time-averaged propulsion force $\vec{F}$ and propulsion torque $T$. 
The propulsion force can be decomposed as $\vec{F}=F_\parallel\uu_\parallel+F_\perp\uu_\perp$ into a component $F_\parallel$ parallel to $\uu_\parallel$ and a component $F_\perp$ parallel to $\uu_\perp$.

\subsection{Parameters}
Table \ref{tab:Parameters} gives an overview of the parameters that are relevant to our study. 
Their values are chosen analogously to Ref.\ \cite{VossW2020}.
\begin{table*}[htb]
\centering
\caption{\label{tab:Parameters}Relevant parameters and their values.}% 
\begin{ruledtabular}
\begin{tabular}{llll}%
\textbf{Name} & \textbf{Symbol} & \textbf{Value} & \textbf{Note}\\
\hline
Subparticle diameter & $\sigma$ & $\SI{1}{\micro\metre}$ &\\
Subparticle height & $h$ & $\sigma/2$ &\\
Overlap width & $w$ & $\sigma/2$ &\\
Particle orientation angle & $\theta$ & $0$-$\pi$ &\\
Sound frequency & $f$ & $\SI{1}{\mega\hertz}$ &\\
Speed of sound & $c_\mathrm{f}$ & $\SI{1484}{\metre\,\second^{-1}}$ & For water at $T_0$ and $p_0$\\
Time period of sound & $\tau=1/f$ & $\SI{1}{\micro\second}$ &\\
Wavelength of sound & $\lambda=c_\mathrm{f}/f$ & $\SI{1.484}{\milli\metre}$ &\\
Temperature of fluid & $T_0$ & $\SI{293.15}{\kelvin}$ & Standard temperature\\
Mean mass density of fluid & $\rho_0$ & $\SI{998}{\kilogram\,\metre^{-3}}$ & For water at $T_0$ and $p_0$\\
Mean pressure of fluid & $p_{0}$ & $\SI{101325}{\pascal}$ & Standard pressure\\
Initial velocity of fluid & $\ww_{0}$ & $\vec{0}\,\SI{}{\metre\,\second^{-1}}$ &\\
Sound pressure amplitude & $\Delta p$ & $\SI{10}{\kilo\pascal}$ &\\
Acoustic energy density & $E=\Delta p^2/(2 \rho_0 c_{\mathrm{f}}^2)$ & $\SI{22.7}{\milli\joule\,m^{-3}}$ &\\
Shear/dynamic viscosity of fluid & $\nu_{\mathrm{s}}$ & $\SI{1.002}{\milli\pascal\,\second}$ & For water at $T_0$ and $p_0$\\
Bulk/volume viscosity of fluid & $\nu_{\mathrm{b}}$ & $\SI{2.87}{\milli\pascal\,\second}$ & For water at $T_0$ and $p_0$\\
Inlet-particle or particle-outlet distance & $l_1$ & $\lambda/4$ &\\
Inlet length & $l_2$ & $\SI{200}{\micro\metre}$ &\\
Mesh-cell size & $\Delta x$ & $\SI{15}{\nano \metre}$-$\SI{1}{\micro \metre}$ &\\
Time-step size & $\Delta t$ & $1$-$\SI{10}{\pico \second}$ &\\
Simulation duration & $t_{\mathrm{max}}$ & $\geqslant 500\tau$ &\\
Euler number & $\mathrm{Eu}$ &  $\SI{2.2}{\cdot10^5}$ &\\
Helmholtz number & $\mathrm{He}$ & $\SI{1.01}{\cdot10^{-3}}$ &\\
Bulk Reynolds number &  $\mathrm{Re}_\mathrm{b}$ & $\SI{3.52}{\cdot10^{-3}}$ &\\
Shear Reynolds number &  $\mathrm{Re}_\mathrm{s}$ & $\SI{1.01}{\cdot10^{-2}}$ &\\
Particle Reynolds number &  $\mathrm{Re}_\mathrm{p}$ & $<\SI{4}{\cdot10^{-8}}$ &
\end{tabular}%
\end{ruledtabular}%
\end{table*}

\subsection{Acoustofluidic simulations}
We simulate the propagation of a planar traveling ultrasound wave through the considered system and its interaction with the particle by numerically solving the basic equations of fluid dynamics with the finite volume method. 
These equations are the continuity equation describing the time evolution of the mass-density field of the fluid, the compressible Navier-Stokes equations describing the time evolution of the velocity field of the fluid, and a linear constitutive equation for the pressure field of the fluid that is needed as closure for the system of coupled partial differential equations. The implementation of these direct fluid dynamics simulations is based on the finite volume software package OpenFOAM \cite{WellerTJF1998}. 

Nondimensionalization of these equations yields the Euler number $\mathrm{Eu}$, Helmholtz number $\mathrm{He}$, bulk Reynolds number $\mathrm{Re}_\mathrm{b}$, and shear Reynolds number $\mathrm{Re}_\mathrm{s}$ as dimensionless numbers (see Ref.\ \cite{VossW2021} for a more detailed discussion).
Choosing the particle length $2\sigma-w=3\sigma/2$ as characteristic length, the dimensionless numbers are given by 
\begin{align}
\mathrm{Eu}&=\frac{\Delta p}{\rho_0 \Delta u^2}\approx \SI{2.20}{\cdot10^5},\\
\mathrm{He}&=\frac{3 f\sigma}{2 c_\mathrm{f}}\approx \SI{1.01}{\cdot10^{-3}},\\
\mathrm{Re}_\mathrm{b}&=\frac{3 \rho_0 \Delta u \sigma}{2 \nu_\mathrm{b}}\approx \SI{3.52}{\cdot10^{-3}},\\
\mathrm{Re}_\mathrm{s}&=\frac{3 \rho_0 \Delta u \sigma}{2 \nu_\mathrm{s}}\approx \SI{1.01}{\cdot10^{-2}}.
\end{align}
For applying the finite volume method, we discretize the fluid domain by a structured, mixed rectangle-triangle mesh with about 250,000 cells.
The typical cell size $\Delta x$ is $\SI{15}{\nano\metre}$ near the particle and $\SI{1}{\micro\metre}$ far away from it.
We realize the time integration by an adaptive time-step method, where the time-step size $\Delta t = 1\text{-}\SI{10}{\pico \second}$ always meets the Courant-Friedrichs-Lewy condition 
\begin{align}
C = c_\mathrm{f} \frac{\Delta t}{\Delta x} < 1 .
\end{align}
The simulations are run from start time $t=0$ to end time $t_\mathrm{max} \geqslant 500\tau$ with the time period of the ultrasound $\tau=1/f=\SI{1}{\micro\second}$. 
An individual simulation run requires about $36,000$ CPU core hours.

\subsection{Propulsion force and torque}
After performing the acoustofluidic simulations, the time evolution of the mass-density field, velocity field, and pressure field in the system are known.  
From these fields, we then calculate the time-dependent propulsion force and torque that are exerted on the particle in the laboratory frame. 
For this purpose, we calculate the contributions \cite{LandauL1987}
{\allowdisplaybreaks\begin{align}%
F^{(\alpha)}_{i} &= \sum^{2}_{j=1} \int_{\partial\Omega_{\mathrm{p}}} \!\!\!\!\!\!\! \Sigma^{(\alpha)}_{ij}\,\dif A_{j}, \label{eq:F}\\
T^{(\alpha)} &= \sum^{2}_{j,k,l=1} \int_{\partial\Omega_{\mathrm{p}}} \!\!\!\!\!\!\! \epsilon_{ijk}(x_j-x_{\mathrm{p},j})\Sigma^{(\alpha)}_{kl}\,\dif A_{l}
\label{eq:T}%
\end{align}}%
with $\alpha\in\{p,v\}$ to the propulsion force and torque. Here, a superscript \ZT{$(p)$} denotes a pressure contribution and a superscript \ZT{$(v)$} denotes a viscous contribution. 
These contributions correspond to the pressure contribution $\Sigma^{(p)}$ and viscous contribution $\Sigma^{(v)}$ of the fluid's stress tensor $\Sigma=\Sigma^{(p)}+\Sigma^{(v)}$. 
The time-dependent propulsion force and torque are then given by $\vec{F}^{(p)}+\vec{F}^{(v)}$ and $T^{(p)}+T^{(v)}$, respectively.
Furthermore, $\dif\vec{A}(\vec{x})=(\dif A_{1}(\vec{x}),\dif A_{2}(\vec{x}))^{\mathrm{T}}$ denotes the normal and outwards oriented element of the particle's surface $\partial\Omega_{\mathrm{p}}$ at position $\vec{x}\in\partial\Omega_{\mathrm{p}}$, $\epsilon_{ijk}$ is the Levi-Civita symbol, and $\vec{x}_\mathrm{p}$ is the position of $\mathrm{S}$.
During an individual simulation, the particle is held in its position and orientation. 

The time-dependent propulsion force and torque are averaged over one period $\tau$ for large times $t$ and we extrapolate $t \to \infty$ with the extrapolation procedure from Ref.\ \cite{VossW2020}. 
This yields results for the time-averaged propulsion force $\vec{F}=\vec{F}_p+\vec{F}_v$ and torque $T=T_p+T_v$ in the stationary state. 
Their contributions are given by $\vec{F}_p=\langle\vec{F}^{(p)}\rangle$, $\vec{F}_v=\langle\vec{F}^{(v)}\rangle$, $T_p=\langle T^{(p)}\rangle$, and $T_v=\langle T^{(v)}\rangle$ with the time average $\langle\cdot\rangle$. 
The components $F_{\parallel}$ and $F_\perp$ of the time-averaged propulsion force, which are parallel and perpendicular to the unit vector $\uu_\parallel$ describing the particle's orientation, respectively, can be obtained by the projection
\begin{align}
F_{\parallel} &= \vec{F}\cdot\uu_\parallel , \label{eq:parallel_force} \\
F_\perp &= \vec{F}\cdot\uu_\perp . \label{eq:perpendicular_force}
\end{align}

\subsection{Translational and angular propulsion velocity}
From the time-averaged propulsion force components $F_{\parallel}$ and $F_\perp$ and the time-averaged propulsion torque $T$, we calculate the corresponding translational velocities $v_\parallel$ and $v_\perp$ and angular velocity $\omega$.
For this purpose, we apply the Stokes law \cite{HappelB1991}
\begin{equation}
\vec{\mathfrak{v}}=\frac{1}{\nu_\mathrm{s}}\boldsymbol{\mathrm{H}}^{-1}\,\vec{\mathfrak{F}}
\label{eq:velocity}%
\end{equation}
with the translational-angular velocity vector $\vec{\mathfrak{v}}=(v_\parallel,v_\perp,0,0,0,\omega)^{\mathrm{T}}$, force-torque vector $\vec{\mathfrak{F}}=(F_\parallel,F_\perp,0,0,0,T)^{\mathrm{T}}$, shear viscosity of the fluid $\nu_\mathrm{s}$, and hydrodynamic resistance matrix of the particle
\begin{equation}
\boldsymbol{\mathrm{H}}=
\begin{pmatrix}
\boldsymbol{\mathrm{K}} & \boldsymbol{\mathrm{C}}^{\mathrm{T}}_{\mathrm{S}} \\
\boldsymbol{\mathrm{C}}_{\mathrm{S}} & \boldsymbol{\Omega}_{\mathrm{S}} 
\end{pmatrix}.
\label{eq:H}%
\end{equation}
Here, $\boldsymbol{\mathrm{K}}$, $\boldsymbol{\mathrm{C}}_{\mathrm{S}}$, and $\boldsymbol{\Omega}_{\mathrm{S}}$ are submatrices.
The latter two submatrices depend on a reference point that is chosen here to be the center of mass $\mathrm{S}$, as indicated by a subscript $\mathrm{S}$. 

We calculate the values of the submatrices with the software \texttt{HydResMat} \cite{VossW2018,VossJW2019}.
For the particle orientation $\theta=0$, this yields 
\begin{align}
\boldsymbol{\mathrm{K}} &= \begin{pmatrix}
\SI{12.31}{\micro\metre} & \SI{0.11}{\micro\metre} & 0\\
\SI{0.11}{\micro\metre} & \SI{10.60}{\micro\metre} & 0 \\
0 & 0 & \SI{10.79}{\micro\metre}
\end{pmatrix},
\label{eq:K}
\\
\boldsymbol{\mathrm{C}}_{\mathrm{S}} &= \boldsymbol{0},
\\
\boldsymbol{\Omega}_{\mathrm{S}} &= \begin{pmatrix}
\SI{6.63}{\micro\metre^3} & \SI{0.81}{\micro\metre^3} & 0\\
\SI{0.81}{\micro\metre^3} & \SI{5.10}{\micro\metre^3} & 0 \\
0 & 0 & \SI{6.48}{\micro\metre^3}
\end{pmatrix}.
\label{eq:Omega}%
\end{align}
Matrices for other orientations of the particle can be calculated from Eqs.\ \eqref{eq:K}-\eqref{eq:Omega} by a simple transformation that is explicitly stated, e.g., in Ref.\ \cite{VossW2018}.
Since the matrices \eqref{eq:H}-\eqref{eq:Omega} correspond to three spatial dimensions, whereas our acoustofluidic simulations are performed in two spatial dimensions to keep the computational costs affordable, we assume that the particle has a thickness of $\sigma$ in the third dimension when calculating the hydrodynamic resistance matrix $\boldsymbol{\mathrm{H}}$ and we use the three-dimensional versions of Eqs.\ \eqref{eq:F}-\eqref{eq:velocity}.

When the values of $v_\parallel$ and $v_\perp$ are known, one can calculate the particle Reynolds number
\begin{equation}
\mathrm{Re}_\mathrm{p} = \frac{3 \rho_0\sigma}{2 \nu_\mathrm{s}} \sqrt{v_\parallel^2+v_\perp^2}
<\SI{4}{\cdot10^{-8}}.
\end{equation}
Its small value shows that inertial forces corresponding to the particle's motion are dominated by viscous forces.

\section{\label{results}Results and discussion}
Here, we discuss our simulation results for the time-averaged stationary flow field that is generated around the particle depicted in Fig.\ \ref{fig:fig1} and the strength of the associated time-averaged stationary propulsion of the particle.

\subsection{Orientation-dependent flow field}
\begin{figure*}[htb]
\centering
\includegraphics[width=\linewidth]{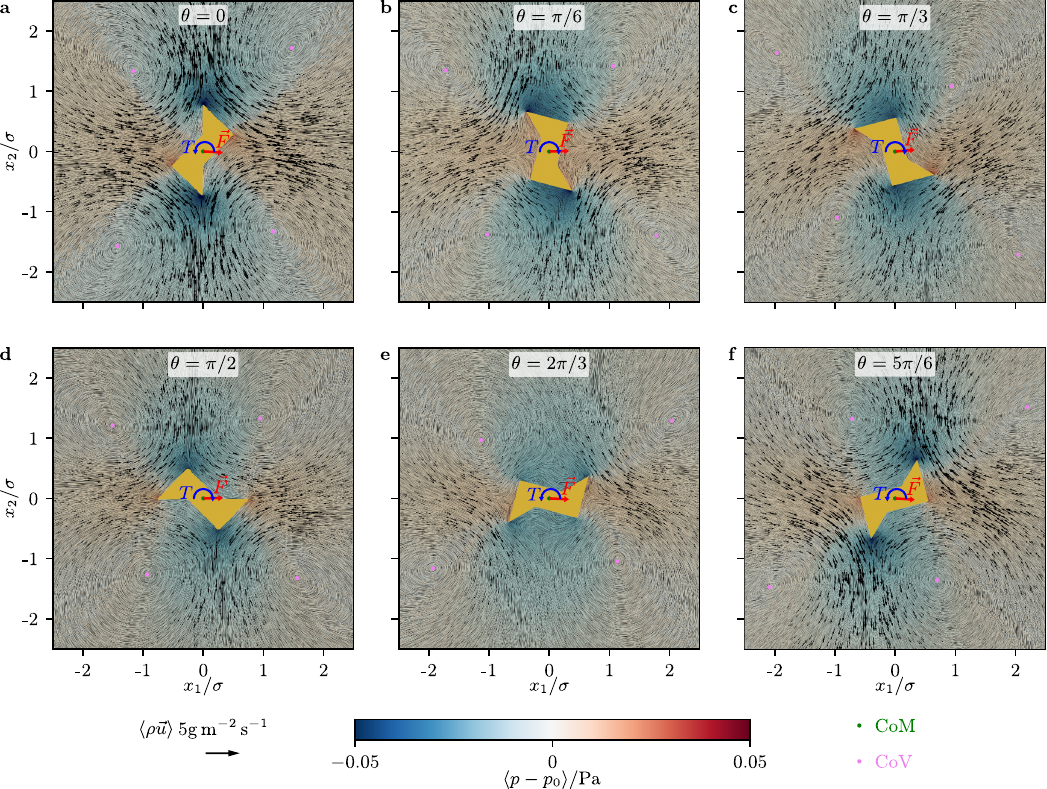}%
\caption{\label{fig:fig2}Time-averaged stationary mass-current density $\langle\rho\ww\rangle$ and reduced pressure $\langle p-p_{0}\rangle$ for varying orientation $\theta$ of the particle from Fig.\ \ref{fig:fig1}.
The center of mass (CoM) of the particle, the centers of vortices (CoV) of the flow field, and the orientations of the particle's propulsion force $\vec{F}$ and torque $T$ are indicated.}
\end{figure*}

We start by discussing how the flow field around the particle depends on the particle's orientation.  
The results of our simulations for the flow field are shown in Fig.\ \ref{fig:fig2}.
As one can see there, the orientation of the particle has only a moderate influence on the flow field. 
Its overall structure is the same for all orientations and similar to the structure of the flow field that has been found for other acoustically propelled particles \cite{VossW2020,VossW2021,VossW2022orientation,VossW2022acoustica,VossW2022acousticb}.
There are 4 vortices at the top left, top right, bottom left, and bottom right of the particle that dominate the flow field. 
As a consequence of these vortices, the fluid flows away from the particle above and below it and towards the particle from the left and right.
Associated with this, the pressure is decreased above and below the particle and increased laterally. 
These positions refer to the laboratory frame and do not rotate with the particle.
This observation is consistent with the orientation dependence of the flow field of a triangular particle that has been published recently \cite{VossW2022orientation}. For different orientations of the particle, we see only small displacements of the centers of the vortices and thus small changes in the structure of the flow field. 
The minima and maxima of the pressure field occur at the tips of the particle and thus rotate with the particle until the tips migrate from a decreased-pressure region to an increased-pressure region or vice versa. 
When a tip with an extremum of the pressure leaves its corresponding region, the extremum switches to another tip in that region.

\subsection{Orientation-dependent propulsion}
\begin{figure*}[htb]
\centering
\includegraphics[width=\linewidth]{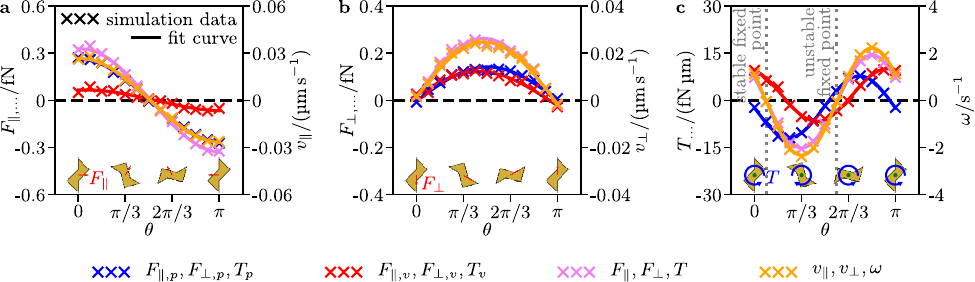}%
\caption{\label{fig:fig3}Simulation data and fit curves for the time-averaged stationary propulsion forces $F_\parallel$ and $F_\perp$ and torque $T$ acting on the particle, their pressure components $F_{\parallel,p}$, $F_{\perp,p}$, and $T_p$, their viscous components $F_{\parallel,v}$, $F_{\perp,v}$, and $T_v$, and the corresponding velocities $v_\parallel$, $v_\perp$, and $\omega$ as functions of the particle's orientation $\theta$. In \textbf{c} also the fixed points of the particle's orientation are indicated.}
\end{figure*}

We proceed to discuss the dependence of the particle's propulsion on the particle's orientation.  
This orientation-dependence is found to be strong. 
The results of our simulations for the propulsion are shown in Fig.\ \ref{fig:fig3}.
This figure shows how the propulsion of the particle depends on the particle's orientation $\theta\in[0,\pi]$ relative to the traveling ultrasound wave that supplies the particle with energy.  
The propulsion is here characterized by the time-averaged stationary propulsion forces $F_\parallel$ and $F_\perp$ parallel and perpendicular to the particle's orientation (as defined in Fig.\ \ref{fig:fig1}), respectively, the time-averaged stationary propulsion torque $T$ acting on the particle, their pressure components $F_{\parallel,p}$, $F_{\perp,p}$, and $T_p$, their viscous components $F_{\parallel,v}$, $F_{\perp,v}$, and $T_v$, as well as the translational propulsion velocities $v_\parallel$ and $v_\perp$ and the angular propulsion velocity $\omega$ that correspond to $F_\parallel$, $F_\perp$, and $T$, respectively.

\subsubsection{Description}
First, we focus on the parallel components of the propulsion (see Fig.\ \ref{fig:fig3}a). 
Their functions look rather simple and similar to a cosine function with period $2\pi$.
All functions have a maximum close to $\theta=0$, a zero at about $\theta=\pi/2$, and a minimum near $\theta=\pi$. 
The pressure component $F_{\parallel,p}$ decreases from $F_{\parallel,p}=\SI{2.72}{\cdot10^{-1}\,\femto\newton}$ to $F_{\parallel,p}=\SI{-2.72}{\cdot10^{-1}\,\femto\newton}$, the viscous component $F_{\parallel,v}$ decreases from $F_{\parallel,v}=\SI{5.04}{\cdot10^{-2}\,\femto\newton}$ to $F_{\parallel,v}=\SI{-5.04}{\cdot10^{-2}\,\femto\newton}$,  
the parallel propulsion force $F_\parallel$ decreases from $F_\parallel=\SI{3.23}{\cdot10^{-1}\,\femto\newton}$ to $F_\parallel=\SI{-3.23}{\cdot10^{-1}\,\femto\newton}$, 
and the parallel propulsion velocity $v_\parallel$ decreases from $v_\parallel=\SI{2.62}{\cdot10^{-2}\,\micro\metre\,\second^{-1}}$ to $v_\parallel=\SI{-2.62}{\cdot10^{-2}\,\micro\metre\,\second^{-1}}$.
 
Second, we consider the perpendicular components of the propulsion (see Fig.\ \ref{fig:fig3}b). 
Also, the functions of these components look rather simple. Now, the functions are similar to a sine function with period $2\pi$ and have minima at $\theta=0$ and $\theta=\pi$ and a maximum close to $\theta=\pi/2$.
The pressure component $F_{\perp,p}$ starts with $F_{\perp,p}=\SI{-7.45}{\cdot10^{-3}\,\femto\newton}$, increases to a maximum with $F_{\perp,p}=\SI{1.33}{\cdot10^{-1}\,\femto\newton}$, and then decreases to $F_{\perp,p}=\SI{7.45}{\cdot10^{-3}\,\femto\newton}$, the viscous component $F_{\perp,v}$ starts with $F_{\perp,v}=\SI{2.73}{\cdot10^{-2}\,\femto\newton}$, increases to $F_{\perp,v}=\SI{1.25}{\cdot10^{-1}\,\femto\newton}$, and decreases to $F_{\perp,v}=\SI{-2.73}{\cdot10^{-2}\,\femto\newton}$, the perpendicular propulsion force $F_\perp$ starts with $F_\perp=\SI{1.98}{\cdot10^{-2}\,\femto\newton}$, increases to $F_\perp=\SI{2.57}{\cdot10^{-1}\,\femto\newton}$, and decreases to $F_\perp=\SI{-1.98}{\cdot10^{-2}\,\femto\newton}$, 
and the perpendicular propulsion velocity $v_\perp$ starts with $v_\perp=\SI{1.54}{\cdot10^{-3}\,\micro\metre\,\second^{-1}}$, increases to $v_\perp=\SI{2.41}{\cdot10^{-2}\,\micro\metre\,\second^{-1}}$, and decreases to $v_\perp=\SI{-1.54}{\cdot10^{-3}\,\micro\metre\,\second^{-1}}$.
 
Third, we address the angular components of the propulsion (see Fig.\ \ref{fig:fig3}c). 
Again, the functions of the components look similar to a cosine function, but now it has period $\pi$ and significant phase shifts and offsets are visible. 
The pressure component $T_p$ starts with $T_p=\SI{-2.25}{\femto\newton\,\micro\metre}$, decreases to a minimum $T_p=\SI{-11.95}{\femto\newton\,\micro\metre}$ at $\theta=\pi/4$, increases through zero near $\theta=\pi/2$ to a maximum $T_p=\SI{7.81}{\femto\newton\,\micro\metre}$ at $\theta=3\pi/4$, and decreases again to $T_p=\SI{-2.25}{\femto\newton\,\micro\metre}$.
On the other hand, the viscous component $T_v$ starts with $T_v=\SI{9.56}{\femto\newton\,\micro\metre}$, decreases to a minimum $T_v=\SI{-6.62}{\femto\newton\,\micro\metre}$ at $\theta=5\pi/12$, increases until a maximum $T_v=\SI{9.71}{\femto\newton\,\micro\metre}$ at $\theta=11\pi/12$, and decreases back to $T_v=\SI{9.56}{\femto\newton\,\micro\metre}$.
The propulsion torque $T$ starts with $T=\SI{7.31}{\femto\newton\,\micro\metre}$, decreases to a minimum $T=\SI{-15.46}{\femto\newton\,\micro\metre}$ at $\theta=\pi/3$, increases until a maximum $T=\SI{14.52}{\femto\newton\,\micro\metre}$ at $5\pi/6$, and decreases again to $T=\SI{7.31}{\femto\newton\,\micro\metre}$. 
Finally, the angular propulsion velocity $\omega$ starts with $\omega=\SI{1.13}{\second^{-1}}$, decreases to a minimum $\omega=\SI{-2.38}{\second^{-1}}$ at $\theta=\pi/3$, increases until a maximum $\omega=\SI{2.24}{\second^{-1}}$ at $\theta=5\pi/6$, and decreases back to $\omega=\SI{1.13}{\second^{-1}}$.
 
Thus, the angular propulsion velocity $\omega$ has two zeros for $0\leqslant\theta\leqslant\pi$.
They are approximately at $\theta=\pi/12$ and $\theta=7\pi/12$ and constitute fixed points for the orientation of the particle.
The fixed point near $\theta=\pi/12$ is stable, whereas the fixed point near $\theta=7\pi/12$ is unstable.   
In contrast to particles with simpler shapes that have been studied before, such as triangular particles \cite{VossW2022orientation}, the fixed points of the orientation are now at orientations where the particle is neither exactly parallel nor perpendicular to the propagation direction of the ultrasound wave. 
The occurrence of a stable fixed point means that the proposed particle will not rotate persistently when it is exposed to a planar traveling ultrasound wave.

\subsubsection{\label{sec:analyticrepresentation}Analytic representation}
For further analysis (see Section \ref{sec:orientationaverage}) and to make it easier for readers to build upon our results, we here present a simple analytic representation of the observed orientation-dependent propulsion of the proposed particle. 

From the symmetry properties of the system (see Fig.\ \ref{fig:fig1}), we know that all forces and translational velocities, i.e., the quantities $F_{\parallel,p}(\theta)$, $F_{\parallel,v}(\theta)$, $F_\parallel(\theta)$, $v_\parallel(\theta)$, $F_{\perp,p}(\theta)$, $F_{\perp,v}(\theta)$, $F_\perp(\theta)$, and $v_\perp(\theta)$, are periodic functions $\mathfrak{f}(\theta)$ with period $2\pi$ and the property $\mathfrak{f}(\theta+\pi)=-\mathfrak{f}(\theta)$, 
whereas all torques and the angular velocity, i.e., the quantities $T_p(\theta)$, $T_v(\theta)$, $T(\theta)$, and $\omega(\theta)$, are periodic functions $\mathfrak{t}(\theta)$ with period $\pi$.
Furthermore, we know from the simulation results (see Fig.\ \ref{fig:fig3}) that all these functions are rather simple and that the data for the torques and angular velocity involve offsets. 

Therefore, we use the following low-order Fourier series ansatz for an analytic representation of the simulation data:
\begin{align}
\begin{split}
&\mathfrak{f}(\theta) = a_1 \sin(\theta) + a_2 \cos(\theta) \\
&\quad\text{for }\mathfrak{f}\in\{F_{\parallel,p}, F_{\parallel,v}, F_\parallel, v_\parallel, F_{\perp,p}, F_{\perp,v}, F_\perp, v_\perp\},
\end{split}\label{eq:fit_force_rotation_particle}\\
\begin{split}
&\mathfrak{t}(\theta) = a_0 + a_1 \sin(2\theta) + a_2 \cos(2\theta) \\
&\quad\text{for }\mathfrak{t}\in\{T_p, T_v, T, \omega\}.
\end{split}\label{eq:fit_torque_rotation_particle}
\end{align}
Fitting the coefficients of these functions to the corresponding quantities that characterize the propulsion of the proposed particle leads to the fit values that are listed in Tab.\ \ref{tab:fitparameters}. 
\begin{table*}[htb]
\centering
\caption{\label{tab:fitparameters}Coefficients of the functions \eqref{eq:fit_force_rotation_particle} and \eqref{eq:fit_torque_rotation_particle} and their fit values for all considered quantities characterizing the particle's acoustic propulsion.}
\begin{ruledtabular}
\begin{tabular}{llll}
\textbf{Quantity}& $\mathbf{\boldsymbol{a}_{0}}$ & $\mathbf{\boldsymbol{a}_{1}}$ & $\mathbf{\boldsymbol{a}_{2}}$\\
\hline
$F_{\parallel,p}$ & \NichtAnwendbar & $\SI{5.49}{\cdot10^{-3}\,\femto\newton}$ & $\SI{2.62}{\cdot10^{-1}\,\femto\newton}$ \\ 
$F_{\parallel,v}$ & \NichtAnwendbar & $\SI{9.45}{\cdot10^{-3}\,\femto\newton}$ & $\SI{6.59}{\cdot10^{-2}\,\femto\newton}$ \\
$F_{\parallel}$ & \NichtAnwendbar & $\SI{1.49}{\cdot10^{-2}\,\femto\newton}$ & $\SI{3.28}{\cdot10^{-1}\,\femto\newton}$ \\
$v_{\parallel}$ & \NichtAnwendbar & $\SI{9.43}{\cdot10^{-4}\,\micro\metre\,\second^{-1}}$ & $\SI{2.66}{\cdot10^{-2}\,\micro\metre\,\second^{-1}}$ \\
$F_{\perp,p}$ & \NichtAnwendbar & $\SI{1.42}{\cdot10^{-1}\,\femto\newton}$ & $\SI{-2.60}{\cdot10^{-3}\,\femto\newton}$ \\ 
$F_{\perp,v}$ & \NichtAnwendbar & $\SI{1.20}{\cdot10^{-1}\,\femto\newton}$ & $\SI{2.09}{\cdot10^{-2}\,\femto\newton}$ \\
$F_{\perp}$ & \NichtAnwendbar & $\SI{2.62}{\cdot10^{-1}\,\femto\newton} $ & $\SI{1.83}{\cdot10^{-2}\,\femto\newton}$ \\
$v_{\perp}$ & \NichtAnwendbar & $\SI{2.47}{\cdot10^{-2}\,\micro\metre\,\second^{-1}}$ & $ \SI{1.39}{\cdot10^{-3}\,\micro\metre\,\second^{-1}}$ \\
$T_p$ & $\SI{-2.05}{\femto\newton\,\micro\metre}$ & $\SI{-9.82}{\femto\newton\,\micro\metre}$ & $\SI{-3.58}{\cdot10^{-1}\,\femto\newton\,\micro\metre}$\\
$T_v$ & $\SI{1.70}{\femto\newton\,\micro\metre}$ & $\SI{-2.93}{\femto\newton\,\micro\metre}$ & $\SI{7.85}{\femto\newton\,\micro\metre}$\\
$T$ & $\SI{-3.49}{\cdot10^{-1}\,\femto\newton\,\micro\metre}$ & $\SI{-12.75}{\femto\newton\,\micro\metre}$ & $\SI{7.49}{\femto\newton\,\micro\metre}$\\
$\omega$ & $\SI{-5.38}{\cdot10^{-2}\,\second^{-1}}$ & $\SI{-1.96}{\second^{-1}}$ & $\SI{1.15}{\second^{-1}}$
\end{tabular} 
\end{ruledtabular}%
\end{table*}
As can be seen in Fig.\ \ref{fig:fig3}, the agreement of the fit functions with our simulation data is excellent already for this low-order Fourier series ansatz.

\subsection{\label{sec:orientationaverage}Orientation-averaged propulsion}
We now consider the orientation-averaged propulsion of the particle, which is relevant, e.g., to predict the motion of such a particle when it is exposed to isotropic ultrasound.
Because of their excellent agreement with the simulation data, we here use the fit functions for the quantities $F_{\parallel,p}(\theta)$, $F_{\parallel,v}(\theta)$, $F_\parallel(\theta)$, $v_\parallel(\theta)$, $F_{\perp,p}(\theta)$, $F_{\perp,v}(\theta)$, $F_\perp(\theta)$, $v_\perp(\theta)$, $T_p(\theta)$, $T_v(\theta)$, $T(\theta)$, and $\omega(\theta)$ from Section \ref{sec:analyticrepresentation}. 

Averaging these functions over the particle's orientation $\theta\in[0,2\pi)$ yields 
\begin{align}
\langle\mathfrak{f}(\theta)\rangle_\theta &= 0 
\text{ for }\mathfrak{f}\in\{F_{\parallel,p}, F_{\parallel,v}, F_\parallel, v_\parallel, F_{\perp,p}, F_{\perp,v}, F_\perp, v_\perp\},
\label{eq:fit_force_rotation_particle_average}\\
\langle\mathfrak{t}(\theta)\rangle_\theta &= a_0 
\text{ for }\mathfrak{t}\in\{T_p, T_v, T, \omega\},
\label{eq:fit_torque_rotation_particle_average}
\end{align}
where $\langle\cdot\rangle_\theta$ denotes the orientational average. 
This shows that the proposed particle will exhibit no translational but significant rotational propulsion when it is exposed to isotropic ultrasound or a circular standing ultrasound wave \cite{MohantyEtAl2021}. 
The particle will therefore rotate persistently and with constant angular velocity, and can thus be called a \textit{nano- or microspinner}. 
This is an interesting feature of the particle. While the particle considered here exhibits pure rotational propulsion, triangular particles, whose orientation-dependent propulsion was studied earlier \cite{VossW2022orientation}, exhibit pure translational forward propulsion in isotropic ultrasound. 

The orientation-averaged angular propulsion velocity of the particle studied here is $\langle\omega\rangle_\theta=\SI{-5.38}{\cdot10^{-2}\,\second^{-1}}$ for the (rather low) ultrasound intensity that we have chosen in this work.
Since the acoustic propulsion velocity is proportional to the energy density $E$ of the ultrasound \cite{VossW2022acoustica}, 
a faster rotation of the particle can easily be achieved by increasing the ultrasound intensity.

\section{\label{conclusions}Conclusions}
We have proposed and studied a particle design that exhibits orientation-dependent translational and rotational propulsion in a planar traveling ultrasound wave but pure and constant rotational propulsion when it is exposed to isotropic ultrasound. 
In the latter scenario, such nano- and microspinners could be applied as nano- and micromixers for mixing fuels at a microscopic level, where the ultrasound propulsion allows to control the amount of mixing spatially and temporally with a fine resolution.
Compared to previously proposed designs for nano- or microspinners, which have been considered and shown to be functional only in a planar \cite{WangCHM2012,BalkEtAl2014,AhmedGFM2014,ZhouZWW2017,SabrinaTABdlCMB2018,LiuR2020,KaynakONNLCH2016,KaynakONLCH2017} or circular \cite{MohantyEtAl2021} standing ultrasound wave, our particle design has significant advantages. 
In particular, an (approximately) isotropic ultrasound field is easier to realize in large bulk systems than a standing ultrasound wave, and particles in isotropic ultrasound can move freely in space (e.g., by Brownian motion), since isotropic ultrasound has no preferred direction, whereas particles in a standing ultrasound wave are usually trapped in a nodal plane of the ultrasound field \cite{WangCHM2012,BalkEtAl2014,ZhouZWW2017,SabrinaTABdlCMB2018,AhmedEtAl2013,GarciaGradillaEtAl2013,GarciaGradillaSSKYWGW2014,WangLMAHM2014,AhmedGFM2014,WuEtAl2015a,EstebanFernandezdeAvilaMSLRCVMGZW2015,Kiristi2015,SotoWGGGLKACW2016,AhmedWBGHM2016,EstebanEtAl2016,ZhouYWDW2017,EstebanFernandezEtAl2017,UygunEtAl2017,WangGWSGXH2018,EstebanEtAl2018,QualliotineEtAl2019,BhuyanDBSGB2019,DumyJMBGMHA2020,LiuR2020,ValdezLOESSWG2020,McneillSWOLNM2021}.

Future research should place greater emphasis on particles with rotational acoustic propulsion and continue our research, e.g., by studying the proposed particle in more detail or by proposing and studying further designs of nano- and microspinners that are propelled by isotropic ultrasound. 
Regarding the former case, the calculation of actual trajectories (taking, e.g., Brownian motion, externally imposed flow fields in the fluid, and further particles into account) remains an important task for future research. 
The simple analytic expressions for the particle's propulsion that we have presented alongside our simulation results in this work allow to describe the trajectories of the particle by Langevin equations \cite{WittkowskiL2012,tenHagenWTKBL2015} or field theories \cite{BickmannW2020twoD,BickmannW2020b} where the acoustic propulsion is taken into account implicitly through an effective propulsion force and torque. Computationally highly expensive acoustofluidic simulations, as we have performed for the present work, are then no longer necessary.

\section*{Data availability}
The raw data corresponding to the figures shown in this article are available as Supplementary Material \cite{SI}.

\section*{Conflicts of interest}
There are no conflicts of interest to declare.

\begin{acknowledgments}
We thank Patrick Kurzeja for helpful discussions. 
R.W.\ is funded by the Deutsche Forschungsgemeinschaft (DFG, German Research Foundation) -- WI 4170/3-1. 
The simulations for this work were performed on the computer cluster PALMA II of the University of M\"unster. 
\end{acknowledgments}

% \clearpage
\nocite{apsrev41Control}
\bibliographystyle{apsrev4-1}
\bibliography{control,refs}
	
\end{document}